\documentclass[12pt]{iopart}
\usepackage[]{algorithm2e}
\usepackage{amssymb}
\usepackage{dsfont}
\usepackage{tensor}
\usepackage{upgreek}
\usepackage{braket}
\usepackage{upgreek}
\usepackage{tensor}
\usepackage{dsfont}
\usepackage{listings}
\usepackage{graphicx}
\usepackage{hyperref}
\usepackage{matlab-prettifier}

\begin{document}

\title[Coding closed and open quantum systems in MATLAB]{Coding closed and open quantum systems in MATLAB: applications in quantum optics and condensed matter}

\author{Ariel Norambuena$^{1}$, Diego Tancara$^{2}$, and Ra\'ul Coto$^{1}$}

\address{$^1$ Centro de Investigaci\'on DAiTA Lab, Facultad de Estudios Interdisciplinarios, Universidad Mayor, Chile}
\address{$^2$ Faculty of Physics, Pontificia Universidad Cat\'olica de Chile, Avda. Vicu\~{n}a Mackenna 4860, Santiago, Chile}
\ead{ariel.norambuena@umayor.cl}
\vspace{10pt}
\begin{indented}
\item[]
\end{indented}

\begin{abstract}
We develop a package of numerical simulations implemented in MATLAB to solve complex many-body quantum systems. We focus on widely used examples that include the calculation of the magnetization dynamics for the closed and open Ising model, dynamical quantum phase transition in cavity QED arrays, Markovian dynamics for interacting two-level systems, and the non-Markovian dynamics of the pure-dephasing spin-boson model. These examples will be useful for undergraduate and graduate students with a medium or high background in MATLAB, and also for researches interested in numerical studies applied to quantum optics and condensed matter systems.
\end{abstract}

\section{Introduction}
Myriad current research fields in quantum optics and condensed matter demand numerical analysis of complex many-body systems (MBS). Interesting effects like dynamical quantum phase transition (DQPT), meta-stability, steady states, among others, can naturally emerge in MBS. These effects are observed in closed decoherence-free dynamics such as the Jaynes-Cummings~\cite{JaynesCummings1963}, Jaynes-Cummings-Hubbard~\cite{Schmidt2009}, trapped ions~\cite{Cirac2004}, cavity opto-mechanics~\cite{Markus2014} or Ising models~\cite{Dominic2016}. Furthermore, for open quantum systems, the environment may cause memory effects in the reservoir time scale, which is known as non-Markovianity (NM) \cite{Breuer,Vega}. From the theory of open quantum systems, NM can be described either by a time-local convolutionless or a convolution master equation~\cite{Sergey2018}. In most of these cases, it is highly demanding to handle these kinds of problems with analytical tools, and a numerical approach can appear as the only solution. For instance, this is the case for systems with non-trivial system-environment interactions \cite{Ariel2020} and quantum systems with many degrees of freedom. \par

Nowadays, numerical toolboxes or open-source packages allow saving time when dealing with analytically untreatable problems, without the requirement of a vast knowledge in computational physics. Toward this end, we found the Wave Packet open-source package of MATLAB~\cite{Schmidt2017,Schmidt2018,Schmidt2019}, the tutorial Doing Physics with MATLAB~\cite{Cooper}, the introductory book Quantum Mechanics with MATLAB~\cite{Chelikowsky}, and the Quantum Optics Toolbox of MATLAB~\cite{TanQOT}. However, most of them does not have examples illustrating many-body effects in both closed and open quantum systems. In this work, we provide several examples implemented in MATLAB to code both closed and open dynamics in many-body systems. This material will be useful for undergraduate and graduate students, either master or even doctoral courses. More importantly, the advantage of using the matrix operations in MATLAB can be crucial to work with Fock states, spin systems, or atoms coupled to light. \par

MATLAB is a multi-paradigm computational language that provides a robust framework for numerical computing based on matrix operations~\cite{Korsch2016}, along with a friendly coding and high performance calculations. Hence, the solutions that we provide have a pedagogical impact on the way that Quantum Mechanical problems can be efficiently  tackle down. This paper is organized as follows. In section~\ref{CloseQuantumSystems} we introduce the calculation of observables for a decoherence-free closed quantum many-body systems. We discuss the transverse Ising and the Jaynes-Cumming-Hubbard models with a particular interest in dynamical quantum phase transitions. In Section~\ref{OpenQuantumSystems}, we model the Markovian and non-Markovian dynamics of open quantum systems using the density matrix. The dissipative dynamics of interacting two-level systems is calculated using a fast algorithm based on eigenvalues and eigenmatrices. Finally, the non-Markovian dynamics of the pure-dephasing spin-boson model is discussed and modelled for different spectral density functions, in terms of a convolutionless master equation.

\section{Closed quantum systems} \label{CloseQuantumSystems}
In the framework of closed quantum many-body systems the dynamics is ruled by time-dependent Shcr\"{o}dinger equation~\cite{Schrodinger1926} 

\begin{equation}
i\hbar {d |\Psi(t) \rangle \over dt} = \hat{H}|\Psi(t) \rangle ,  
\end{equation}

where $|\Psi(t) \rangle$ and $\hat{H}$ are the many-body wavefunction and Hamiltonian of the system, respectively. For a time-independent Hamiltonian, the formal solution for the wavefunction reads

\begin{equation}\label{ManyBodyWavfunction}
|\Psi(t) \rangle  = \hat{U}(t)|\Psi(0)\rangle = e^{-i \hat{H} t/ \hbar}|\Psi(0)\rangle,
\end{equation}

where $\hat{U}(t) = \mbox{exp}(-i \hat{H} t/ \hbar)$ is the time propagator operator. The numerical implementation of this type of dynamics relies on the construction of the initial state $|\Psi(0)\rangle$ (vector), the time propagator $\hat{U}(t)$ (matrix), and the system Hamiltonian $\hat{H}$ (matrix). For illustration, we introduce first two spin-$1/2$ particles described by the following Hamiltonian

\begin{equation} \label{SpinHamiltonian}
\hat{H}_{\rm spins} = -J S_1^{x} S_2^{x} - B(S_1^{x}+S_2^{x}).
\end{equation}

In MATLAB, the above Hamiltonian can be coded as follow (using $J = 1$ and $B = 0.1J$)

\begin{lstlisting}[basicstyle=\tiny,style=Matlab-editor,basicstyle=\color{black}\ttfamily\scriptsize]
J = 1;              % Value of the coupling J
B = 0.1*J;          % Value of the magnetic field
Sx = [0 1;1 0];     % S_x operator
Sz = [1 0; 0 -1];   % S_z operator
I = eye(2);         % Identity matrix
Hspins = -J*kron(Sx,Sx)-B*(kron(Sx,I)+kron(I,Sx));  % Hamiltonian
\end{lstlisting}

where eye(2) generates the $2 \times 2$ identity matrix and kron($X$,$Y$) = $X \otimes Y$ gives the tensor product of matrices $X$ and $Y$. Now, we illustrate how to solve the spin dynamics for the initial condition $|\Psi(0)\rangle = \ket{\downarrow}_1 \otimes \ket{\downarrow}_2$ from the initial time $t_i = 0$ to the final time $t_f = 2T$ with $T = 2\pi/(2B)$. We must point out that the method we will use next is not valid for time-dependent Hamiltonians. For completeness, we compute the average magnetization along the $\alpha$ direction, which is defined as 

\begin{equation} \label{Magnetization}
\langle M_{\alpha} \rangle = {1 \over N}\sum_{i=1}^{N}\langle \Psi(t)|S_i^{\alpha}|\Psi(t)\rangle, \quad \alpha = x,y,z,
\end{equation}

where $N$ is the number of spins and $|\Psi(t)\rangle$ is the many-body wavefunction given in equation~(\ref{ManyBodyWavfunction}). In our particular case, $N=2$, and thus we can write the following code to compute $\langle M_z \rangle$

\begin{lstlisting}[basicstyle=\tiny,style=Matlab-editor,basicstyle=\color{black}\ttfamily\scriptsize]
down = [0 1]';                   % Quantum state down = [0 1]^T
Psi_0 = kron(down,down);         % Initial wavefunction
T = 2*pi/(2*B);                  % Period of time
Nt = 1000;                       % Number of steps to construct time vector
ti = 0;                          % Initial time
tf = 2*T;                        % Final time
dt = (tf-ti)/(Nt-1);             % Step time dt
t = ti:dt:tf;                    % Time vector
U = expm(-1i*Hspins*dt);         % Time propagator operator U(dt)
SSz = (kron(Sz,I)+kron(I,Sz))/2; % Operator S1^z + S2^z
Mz = zeros(size(t));             % Average magnetization
for n=1:length(t)                % Iteration to find Psi(t) and Mz(t)
    if n==1
        Psi = Psi_0;             % Initial wavefunction
    else
        Psi = U*Psi;             % Wavefuntion at time t_n
    end
    Mz(n) = Psi'*SSz*Psi;        % Average magnetization at time t_n
end
plot(t/T,real(Mz),'r-','LineWidth',3)
xlabel('$t/T$','Interpreter','LaTex','Fontsize', 30)
ylabel('$\langle M_z \rangle$','Interpreter','LaTex','Fontsize', 30)
set(gca,'fontsize',21)
\end{lstlisting}

\begin{figure}[htb]
\centerline{\includegraphics[width=1\textwidth]{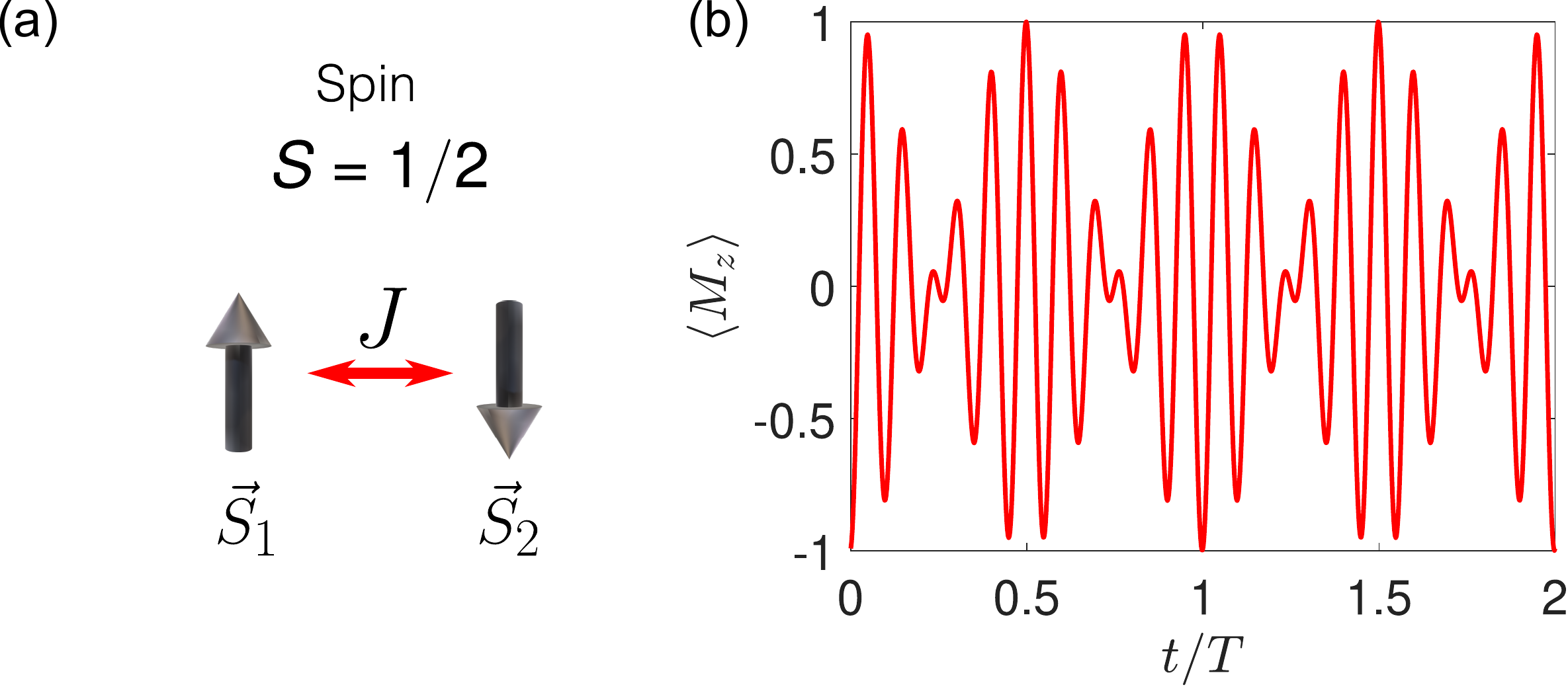}}
\caption{(a) Schematic representation of the two-spin system. (b) Average magnetization along the $z$ direction for $J = 1$ and $B = J/10$ and considering the initial condition$|\Psi(0)\rangle = \ket{\downarrow}_1 \otimes \ket{\downarrow}_2$. The time is divided by the natural period $T = 2\pi/(2B).$} \label{example1}
\end{figure}

By carefully looking at our code, we observed that MATLAB provides an intuitive platform to simulate quantum dynamics. First, the Hamiltonian defined in equation~(\ref{SpinHamiltonian}) can be easily implemented if operators (matrices $S_i^x$) are defined, and the two-body interaction is written using the kron() function. The tensor product defined as kron() reconstructs the Hilbert space of the system (two spin-1/2 particles). Second, the time evolution only depends on the initial state (vector Psi\_0) and the time propagator U = expm(-1i*H*dt), where dt is the time step. We remark that it is convenient to use the expm() function of MATLAB to calculate the exponential of a matrix. Finally, the \textit{for loop} allows us to actualize the wavefunction $|\Psi(t)\rangle$ at every time using the recursive relation Psi = U*Psi ($\Psi_{n+1} = U \Psi_n$), where the term of the left hand is the wavefunction at time $t_{n+1} = t_n + dt$. To have numerical stability is necessary to satisfy the condition $dt \ll \hbar/\max(|\hat{H}|)$. Hence, using the wavefunction we can calculate the average magnetization $\langle M_z \rangle = \langle \Psi(t)| M_z | \Psi(t)\rangle$ employing the command Mz(n) = Psi'*SSz*Psi. \par

In figure~\ref{example1} we plotted the expected average magnetization $\langle M_z \rangle$ for the two-spin system. Initially, the system has a magnetization $\langle M_z \rangle = -1$ due to the condition  $|\Psi(0)\rangle = \ket{\downarrow}_1 \otimes \ket{\downarrow}_2$, and then two characteristic oscillations are observed. The slow and fast oscillations in the signal correspond to the influence of $B$ and $J$, respectively. \par 

We remark that this example does not require any additional package to efficiently run the codes. In the following, all codes may be executed using the available MATLAB library and the functions introduced in each example. In the next subsection, we introduce three relevant many-body problems, namely the transverse Ising, the Rabi-Hubbard and the Jaynes-Cummings-Hubbard models. For each case we write down an algorithm to solve the many-body dynamics.

\subsection{Ising model}
Let us consider the following transverse Ising Hamiltonian~\cite{Cirac2004}

\begin{equation} \label{IsingHamiltonian}
\hat{H}_{\rm Ising} = H_1 + H_0 = -\sum_{i \neq j}^{N} J_{ij} \hat{S}_i^x \hat{S}_j^x  - B \sum_{i=1}^{N} \hat{S}_i^{z},
\end{equation}

where $J_{ij}$ is the coupling matrix, $B$ is the external magnetic field, and $N$ is the number of spins. The spin operators $\hat{S}_i^{\alpha}$ with $\alpha = x,y,z$ and $i=1,...,N$ are the Pauli matrices for $S=1/2$. 

It is important to analyse the required memory to simulate the dynamics of the Ising model. The dimension of the Hilbert space for $N$ spin-1/2 particles is $\mbox{dim}_H = 2^N$, and considering that a double variable requires 8 bytes we found that the memory required (in GB) is given by the expression $\mbox{dim}_H^2 \times 8 \times 10^{-9}$. The scaling $\propto \mbox{dim}_H^2$ is due to the Hermitian matrix structure of the Hamiltonian. For instance, to simulate 14 spins, we need at least 2.2 GB of memory. In what follows, the reader must check the computational cost of each example. \par

The interaction between adjacent spins is modelled using $J_{ij} = |i-j|^{-\alpha}/J$, where $\alpha \geq 0$ and $J = (N-1)^{-1}\sum_{i>j}J_{ij}$~\cite{Zunkovic2016,Halimeh2017}. We set $B = J/0.42$ as it was recently used in Ref~\cite{Jurcevic2017} to study dynamical phase transition in trapped ions. The ground state of the Hamiltonian $H_1$ has a double degeneracy given by $H_1 |\Psi_{\eta}\rangle = E_{\eta}|\Psi_{\eta}\rangle$, where $|\Psi_{\rightarrow}\rangle$ and $|\Psi_{\leftarrow}\rangle$ are the degenerate ground states and $E_{\leftarrow} = E_{\rightarrow}$ the corresponding energies. To study the dynamical quantum phase transition of this system we introduce the following rate function $\Lambda(t)$~\cite{Jurcevic2017} 

\begin{equation} \label{SecondOrderParameter}
\Lambda(t) = \min_{\eta = \rightarrow, \leftarrow}\left[-N^{-1}\ln(P_{\eta}(t)) \right],
\end{equation}

where $P_{\eta}(t) = |\langle \Psi_{\eta}|\Psi(t)\rangle|^2$ is the probability to return to the ground state being $|\Psi(t)\rangle = \mbox{exp}(-i\hat{H}_{\rm Ising} t)|\Psi(0)\rangle$ ($\hbar = 1$). To compute the magnetization vector along the $x$ direction we use equation~(\ref{Magnetization}) with $\alpha = x$. The explicit code in MATLAB is showed below 

\begin{lstlisting}[basicstyle=\tiny,style=Matlab-editor,basicstyle=\color{black}\ttfamily\scriptsize]
Nspins = 6;   % Number of spins
H0 = zeros(2^Nspins,2^Nspins);
H1 = zeros(2^Nspins,2^Nspins);
J = 0;
alpha = 0.2;  % Parameter \alpha
for j=1:Nspins
    for i=j+1:Nspins
        Jij = abs(i-j)^(-alpha);
        J = J + (Nspins-1)^(-1)*Jij;
    end
end
B = J/0.42;   % Magnetic field
Sx = [0 1;1 0];
Sz = [1 0;0 -1];
for i=1:Nspins
    Szi = getSci(Sz,i,Nspins);
    H0 = H0 - B*Szi;       % Hamiltonian for the magnetic field
    for j=1:Nspins
        if i~=j
            Sxi = getSci(Sx,i,Nspins);
            Sxj = getSci(Sx,j,Nspins);
            Vij = abs(i-j)^(-alpha)/J;
            H1 = H1 - Vij*Sxi*Sxj; % Interaction Hamiltonian
        end
    end
end
H = H0 + H1;                % Total hamiltonian
xr = [1 1]'/sqrt(2);        % Single-particle state |Psi_{-->}>
xl = [-1 1]'/sqrt(2);       % Single-particle state |Psi_{<--}>
Xr = xr;
Xl = xl;
for n=1:Nspins-1
   Xr = kron(Xr,xr);        % Many-body state |Psi_{-->}>
   Xl = kron(Xl,xl);        % Many-body state |Psi_{<--}>
end
PSI_0 = Xr;                 % Initial condition
ti = 0;                     % Initial time 
tf = 22;                    % Final time 
Nt = 10000;                 % Number of steps
dt = (tf-ti)/(Nt-1);        % Step time dt
t = ti:dt:tf;               % Time vector
U = expm(-1i*H*dt);         % Time propagator operator U(dt)
Mx = zeros(size(t));        % Average Magnetization <M_x(t)>
Lambda = zeros(size(t));    % Rate function \Lambda(t)
SSx = 0;
for i=1:Nspins
        Sxi = getSci(Sx,i,Nspins);
        SSx = SSx + Sxi/Nspins; % Magnetization operator
end
for n=1:length(t)
    if n==1
        PSI = PSI_0;        % Initial wavefunction
    else
        PSI = U*PSI;        % Wavefunction at time t_n 
    end
    Pr = abs(Xr'*PSI)^2;    % Probability state |Psi_{-->}>
    Pl = abs(Xl'*PSI)^2;    % Probability state |Psi_{<--}>
    Lambda(n) = min(-Nspins^(-1)*log(Pr),-Nspins^(-1)*log(Pl));
    Mx(n) =  PSI'*SSx*PSI;  % Average magnetization along x-axis
end

figure()
plot(B*t,Lambda,'b-','Linewidth',3)
xlabel('$B t$','Interpreter','LaTex','Fontsize', 30)
ylabel('$\Lambda(t)$','Interpreter','LaTex','Fontsize', 30)
set(gca,'fontsize',21)
xlim([0 5])

figure()
box on
plot(t*B,real(Mx),'r-','Linewidth',2)
xlabel('$B t$','Interpreter','LaTex','Fontsize', 30)
ylabel('$\langle M_x\rangle$','Interpreter','LaTex','Fontsize', 30)
set(gca,'fontsize',21)
xlim([0 100])
\end{lstlisting}

In the above code, we have introduced the many-body operators Szi, Sxi, and Sxj (see lines 16, 20 and 21). These operators are mathematically given by

\begin{equation}
\hat{S}_i^{\alpha} = \underbrace{\mathds{1} \otimes \mathds{1}\otimes ... \mathds{1}\otimes}_{i-1 \; \mbox{terms}} \hat{S}^{\alpha} \otimes \mathds{1} \otimes .. \otimes \mathds{1},  \quad \quad \alpha = x,y,z,
\end{equation} 

where $\mathds{1}$ is the single-particle identity matrix and $\hat{S}_i^{\alpha}$ are the Pauli matrices introduced in Hamiltonian~(\ref{IsingHamiltonian}). The numerical implementation of the operators $\hat{S}_i^{\alpha}$ is performed using the function getSci(), which is defined as

\begin{lstlisting}[basicstyle=\tiny,style=Matlab-editor,basicstyle=\color{black}\ttfamily\scriptsize]
function Sci = getSci(sc,i,Nspins)
Is = eye(2);
Op_total = cell(1,Nspins);
for site = 1:Nspins
    Op_total{site} = Is+double(eq(i,site))*(sc-Is);
end
Sci = Op_total{1};
for site = 2:Nspins
    Sci = kron(Sci,Op_total{site});
end
end
\end{lstlisting}

\begin{figure}[htb]
\centerline{\includegraphics[width=1\textwidth]{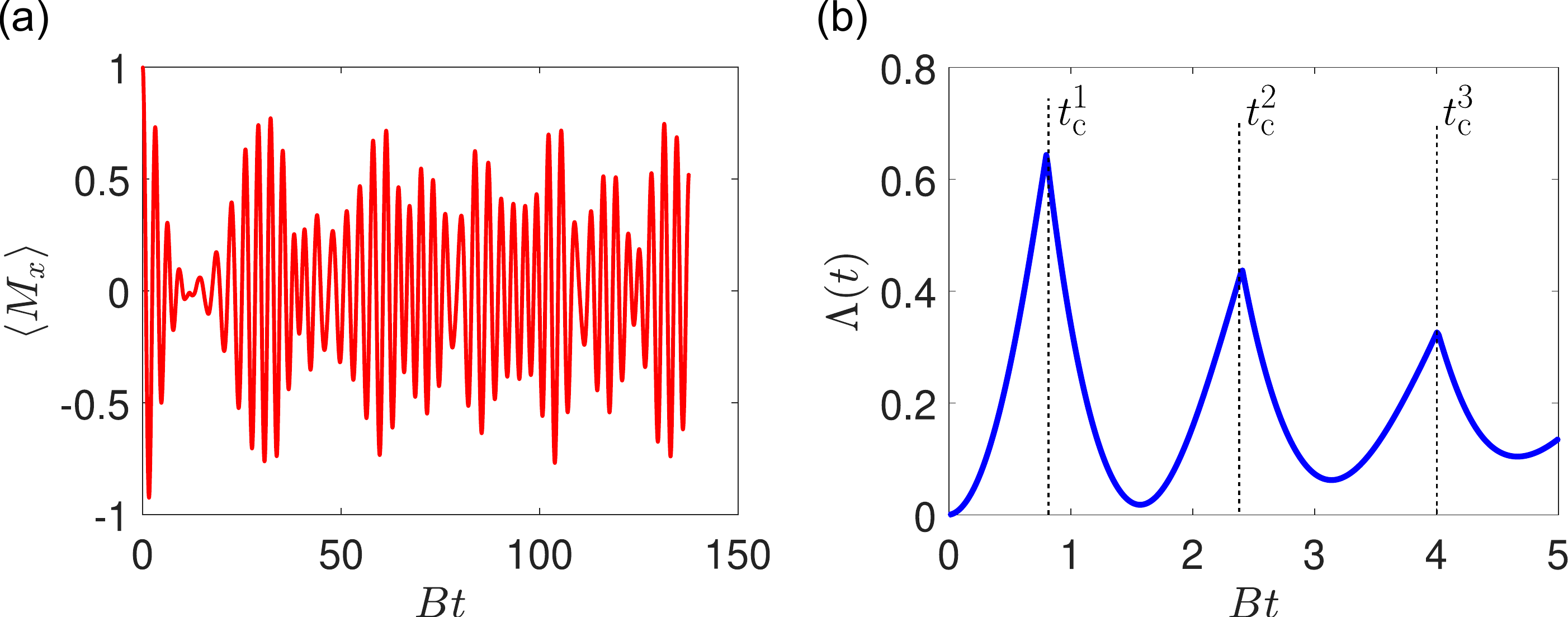}}
\caption{(a) Average magnetization component $\langle M_x \rangle$ for $J/B = 0.42$. (b) Order parameter $\Lambda(t)$ given in equation~(\ref{SecondOrderParameter}) as function of time. The critical times $t_{\rm c}^{i}$ ($i=1,2,3$) are showed in the regions where a non-analytic signature of $\Lambda(t)$ is observed.} \label{figure2}
\end{figure}

In the full code used to study dynamical phase transition in the Ising model we have used the function expm() to calculate the time propagator, similarly to the example previously discussed in Section 2. We remark that the function expm() is calculated using the scaling and squaring algorithm with a Pade approximation (read example 11 of Ref.~\cite{Moler2003} for further details). As a consequence, running the code for high-dimensional matrices, \textit{i.e.}, high-dimensional Hilbert space, increases the consuming time. Therefore, for any many-body Hamiltonian presented in this work, part of the time will depend on the size of the Hilbert space. For instance, for a simulation with $N = 5$ and $N = 10$ spins the presented code requires approximately 0.4 and 120 seconds, respectively, using a computer with 8GB RAM and an Intel core i7 8th gen processor. To profile a MATLAB code we strongly recommend to click on the \textit{Run and Time} bottom to improve the performance of each line in the main code.\par

In figure~\ref{figure2} we plotted the average magnetization $\langle M_x \rangle$ and rate parameter $\Lambda(t)$ for the transverse Ising model. Interestingly, the non-analytic shape of $\Lambda(t)$ is recovered as pointed out in Ref.~\cite{Jurcevic2017}. This non-analytic behaviour prevails at different critical times $t_{\rm c}^{i}$ for which $\left.d \Lambda / dt \right|_{t = t_{\rm c}^i}$ is not well defined. In this example, when $B=0$, the minimum energy is described by the degenerate states $\ket{\Psi}_{\eta}$. After applying the magnetic field $B = 0.42 J$, the system tries to find these minimum energy states. In particular, the sharp peaks in the rate parameter $\Lambda(t)$ show that system switches from the many-body state $\ket{\Psi}_{\rightarrow}$ to $\ket{\Psi}_{\leftarrow}$.

\subsection{Quantum phase transition in cavity QED arrays}
Now, we consider the calculation of a quantum phase transition (QPT) in cavity QED arrays. The dynamics of a system composed by $L$ interacting cavities is described by the Rabi-Hubbard (RH) Hamiltonian~\cite{FelipeTorres2018}

\begin{figure}[htb]
\centerline{\includegraphics[width=1\textwidth]{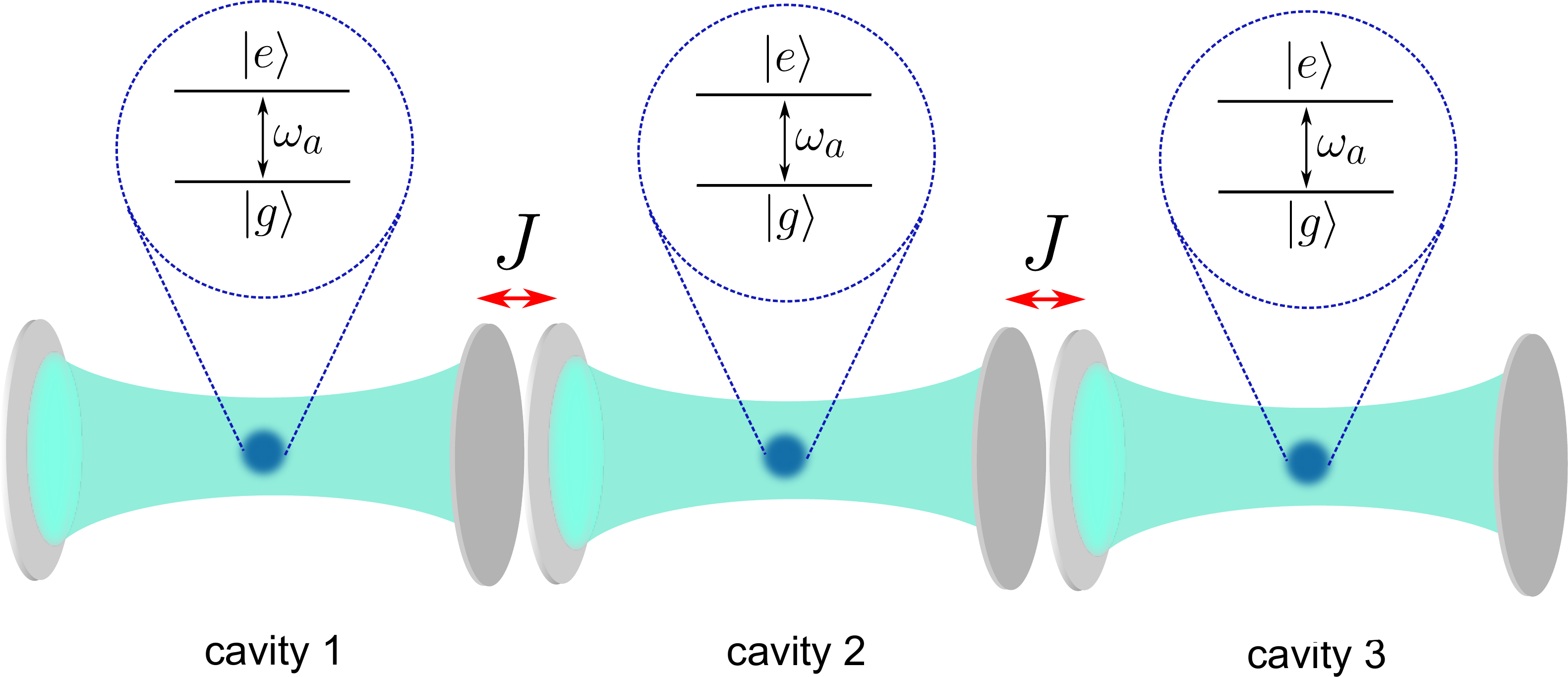}}
\caption{Representation of a cavity QED array with a linear system of three interacting cavities. Inside each cavity we have a two-level system (atom) interacting with a cavity mode. The coupling between adjacent cavities is given by $J$. In the above array, the non-zero elements of the adjacent matrix are $A(1,2) = 1$ and $A(2,3)=1$.} \label{CavityQED}
\end{figure}

\begin{equation}
\hat{H}_{\rm RH} = \sum_{i=1}^{L}\left[\omega_{\rm c} \hat{a}_i^{\dagger}\hat{a}_i + \omega_{\rm a} 
\hat{\sigma}_{i}^{+}\hat{\sigma}_{i}^{-} + g\hat{\sigma}_i^x\left(\hat{a}_i + \hat{a}_i^{\dagger}\right) \right] - J \sum_{i,j}A_{ij} \left( \hat{a}_i \hat{a}_j^{\dagger} + \hat{a}_i^{\dagger} \hat{a}_j\right),
\end{equation}

where we have considered all cavities to be equal, $\omega_{\rm c}$ is the cavity frequency, $\omega_{\rm a}$ is the atom frequency, $g$ is the light-atom coupling constant, $J$ is the photon hopping between neighbouring  cavities, and $A_{ij}$ is the adjacency matrix which takes the values $A_{ij} = 1$ if the cavities are connected, and $A_{ij}=0$ otherwise. In figure~\ref{CavityQED} we show a representation of the system for a linear array of three cavities. \par 

The operators acting on the two-level systems are defined as $\sigma_i^{x} = \sigma_i^{+}+\sigma_i^{-}$, where $\sigma_i^{+} = |e\rangle_i\langle g|_i$ and $\sigma_i^{-} = |g\rangle_i\langle e|_i$, being $\ket{e}_i$ and $\ket{g}_i$ the excited and ground states at site $i$. Note that we are not assuming the first rotating wave approximation (RWA) that leads to $\hat{\sigma}_i^x(\hat{a}_i + \hat{a}_i^{\dagger}) \approx \hat{a}_i \hat{\sigma}_i^+ + \hat{a}_i^{\dagger}\hat{\sigma}_i^-$. In the RWA the fast oscillating terms $\hat{a}_i \hat{\sigma}_i^-$ and $\hat{a}_i^{\dagger}\hat{\sigma}_i^+$ are neglected when the light-atom coupling is small, \textit{i.e.} $g \ll  \omega_{\rm c}$. To study the problem we introduce two relevant Hamiltonians, namely 

\begin{eqnarray}
\hat{H}_{\rm R} &=& \sum_{i=1}^{L}\left[\omega_{\rm c} \hat{a}_i^{\dagger}\hat{a}_i + \omega_{\rm a} 
\hat{\sigma}_{i}^{+}\hat{\sigma}_{i}^{-} + g\hat{\sigma}_i^x\left(\hat{a}_i + \hat{a}_i^{\dagger}\right) \right], \label{Rabi} \\
\hat{H}_{\rm JC} &=& \sum_{i=1}^{L}\left[\omega_{\rm c} \hat{a}_i^{\dagger}\hat{a}_i + \omega_{\rm a} 
\hat{\sigma}_{i}^{+}\hat{\sigma}_{i}^{-} + g\left(\hat{a}_i \hat{\sigma}_i^+ + \hat{a}_i^{\dagger}\hat{\sigma}_i^-\right) \right], \label{JC}
\end{eqnarray}

where equations (\ref{Rabi}) and (\ref{JC}) are the Rabi and Jaynes-Cummings Hamiltonians, respectively. The photon operators acts as follow

\begin{eqnarray}
\hat{a}_i \ket{n_i} = \sqrt{n_i}\ket{n_i-1}, \quad \hat{a}_i^{\dagger} \ket{n_i} = \sqrt{n_i+1}\ket{n_i+1}, 
\end{eqnarray}

where $\ket{n_i}$ is the Fock basis of the $i$-th cavity with $n_i = 0,1,2,...,N_i$, being $N_i$ a cut-off parameter for the Hilbert space of the cavity mode. The many-body wave function at time $t$ can be obtained by propagating the initial condition $\ket{\Psi(0)}$ with the evolution operator $\hat{U}=\mbox{exp}\left( -i \hat{H}_{\rm RH} t\right)$ as follow ($\hbar = 1$)

\begin{equation}
\ket{\Psi(t)} = \mbox{exp}\left( -i \hat{H}_{\rm RH} t\right) \ket{\Psi(0)}, 
\end{equation}

where $\ket{\Psi(0)}$ is the many-body initial state of the system. In this particular case, we choose the Mott-insulator initial condition $\ket{\Psi(0)} = |1,-\rangle_1 \otimes ... \otimes |1,-\rangle_L$, where $|1,-\rangle_i = \cos(\theta_1)\ket{e}_i\otimes\ket{1}_i-\sin(\theta_1)\ket{g}_i\otimes\ket{0}_i$ with $\tan(\theta_n) = 2g\sqrt{n}/\Delta$ being $\Delta = \omega_{\rm a}-\omega_{\rm c}$ the detuning. The quantum phase transition from Mott-insulator to Superfluid can be studied in terms of the following order parameter~\cite{FelipeTorres2018}

\begin{equation} \label{OrderParameter}
\mbox{OP} = {1 \over T}\sum_{i=1}^{L}\int_{0}^{T}\left( \langle \hat{n}_i^2 \rangle- \langle \hat{n}_i \rangle^2  \right) \, d\tau, 
\end{equation}

where $T = J^{-1}$ is an appropriated large time scale to study the dynamics when $J < g < \omega_c$, $\hat{n}_i = \hat{a}_i^{\dagger}\hat{a}_i + \hat{\sigma}_{ee}$ is the number of polaritonic excitations at site $i$, and the expectation values are calculated as $\langle \hat{n}_i^k \rangle= \bra{\Psi(\tau)} \hat{n}_i^k \ket{\Psi(\tau)}$. For comparison we include the rate parameter recently introduced in the Ising model~\cite{Jurcevic2017}

\begin{equation} \label{RateFunction}
\Lambda(t) = - {1 \over L}\mbox{log}_2(P_{1-}(t)).
\end{equation}

In the context of cavity QED arrays $L$ is the number of cavities and $P_{1-}(t) = |\langle \Psi(0) | \Psi(t) \rangle|^2$ is the probability to return to the mott-insulator state $\ket{\Psi(0)} = |1,-\rangle_1 \otimes ... \otimes |1,-\rangle_L$. The following code shows the MATLAB implementation to study this QPT.
\begin{lstlisting}[basicstyle=\tiny,style=Matlab-editor,basicstyle=\color{black}\ttfamily\scriptsize]
Nsim = 25;                      % Number of simulations 
L = 2;                          % Number of cavities
wc = 1;                         % Cavity frequency
g = 1e-2*wc;                    % Atom-light coupling
J = 1e-4*wc;                    % Coupling between cavities
Nph = 2;                        % Number of photons per cavity
dimFock = Nph+1;                % Dimension Fock space for photons
dimT = 2*dimFock;               % Dimension atom+cavity system
Deltai = 10^(-2)*g;             % Initial detuning
Deltaf = 10^(+2)*g;             % Final detuning
xi = log10(Deltai/g);           % Initial detuning in log scale
xf = log10(Deltaf/g);           % Final detuning in log scale
dx = (xf-xi)/(Nsim-1);          % Step dx 
x = xi:dx:xf;                   % Vector x to plot transition phase
OP_JC = zeros(size(x));         % Order parameter Jaynes-Cummings model
OP_R = zeros(size(x));          % Order parameter Rabi model
A = cell(1,L);                  % Cell array to storage \hat{a}_i operators
Sp = cell(1,L);                 % Cell array to storage \hat{\sigma}_i^+ operators
N_ex = cell(1,L);               % Cell array to storage N_i operators
Iatom = eye(2);                 % Identity matrix atom system
Icav = eye(dimFock );           % Identity matrix cavity system
Is = eye(2*dimFock);            % Identity matrix atom+cavity system
for i=1:L
    A{i} = acav(i,L,Nph,Is,Iatom);    % Photon operator \hat{a}_i
    Sp{i} = sigmap(i,L,Is,Icav);       % Atom operator \hat{\sigma}_i 
    N_ex{i} = A{i}'*A{i}+Sp{i}*Sp{i}'; % number of excitations per cavity
end  
Ad = ones(L);                   % Adyacent matrix
Ad = triu(Ad)-eye(L);           % A(i,j)=1 for j>i
Hhopp = zeros(dimT^L,dimT^L);   % Interaction Hamiltonian
for i=1:L
    for j=1:L
        Hhopp = Hhopp - J*Ad(i,j)*A{i}'*A{j} - J*Ad(i,j)*A{i}*A{j}';
    end
end
Nt = 10000;                     % Length time vector 
ti = 0.01/J;                    % Initial time
tf = 1/J;                       % Final time
dt = (tf-ti)/(Nt-1);            % Step time dt
t = ti:dt:tf;                   % Time vector 
Lambda_R = zeros(Nsim,Nt);
Lambda_JC = zeros(Nsim,Nt);
parfor n=1:Nsim
    D = g*10^(x(n));            % Detuning in each simulation
    Model = 'Rabi';             
    [OP_R(n), P1m_R] = QuantumSimulationCavityArray(wc,D,g,Nph,L,A,Sp,N_ex,Hhopp,t,Model);
    Model = 'Jaynes-Cummings';  
    [OP_JC(n),P1m_JC] = QuantumSimulationCavityArray(wc,D,g,Nph,L,A,Sp,N_ex,Hhopp,t,Model);
    Lambda_R(n,:) = -1/L*log2(P1m_R);     % Rate function for the Rabi model
    Lambda_JC(n,:) = -1/L*log2(P1m_JC);   % Rate function for the Jaynes-Cummings model
end  
figure()
box on
hold on
plot(J*t,Lambda_R(end,:),'b-','Linewidth',3)
plot(J*t,Lambda_JC(end,:),'r-','Linewidth',3)
hold off
xlabel('$Jt$','Interpreter','LaTex','Fontsize', 30)
ylabel('$\Lambda(t)$','Interpreter','LaTex','Fontsize', 30)
set(gca,'fontsize',21)
legend({'$\mbox{RH}$','$\mbox{JCH}$'},'Interpreter','latex','Fontsize', 21,'Location','best')

figure()
box on
hold on
plot(x,real(OP_R),'.b','Markersize',30)
plot(x,real(OP_JC),'.r','Markersize',30)
hold off
xlabel('$\mbox{Log}_{10}(\Delta/g)$','Interpreter','LaTex','Fontsize', 30)
ylabel('$\mbox{OP}$','Interpreter','LaTex','Fontsize', 30)
set(gca,'fontsize',21)
legend({'$\mbox{RH}$','$\mbox{JCH}$'},'Interpreter','latex','Fontsize', 21,'Location','best')
\end{lstlisting}

In the above code we are running $N_{\rm sim} = 25$ simulations of the Rabi-Hubbard and Jaynes-Cummings-Hubbard models, one for each value of detuning $\Delta=\omega_{\rm a}-\omega_{\rm c}$, and the initial condition $\ket{\Psi(0)}$ corresponds to each $\Delta$. In line 43 of the main code we introduced a parallel calculation of a cycle {\it for} using the MATLAB function {\it parfor}. In the first iteration the starting parallel pool takes a larger time, but in a second execution of the main code the time is greatly reduced. \par

For simplicity, we are only considering two interacting cavities but the main code can be changed in line 3 to introduce more cavities. Furthermore, the topology of the cavity network can be modified by changing the adjacency matrix $A_{ij}$ in lines 28 and 29.  In addition, in lines 24 and 25 we have introduced the functions acav() and sigmap() to construct the many-body operators $\hat{a}_i$  and $\hat{\sigma}_i^{+}$ for a system of $L$ cavities. These operators are mathematically defined as

\begin{equation}
\hat{a}_i = \underbrace{\mathds{1} \otimes \mathds{1}\otimes ... \mathds{1}\otimes}_{i-1 \; \mbox{terms}} \hat{a} \otimes \mathds{1} \otimes .. \otimes \mathds{1}, 
\quad 
\hat{\sigma}_{i}^{+} = \underbrace{\mathds{1} \otimes \mathds{1}\otimes ... \mathds{1}\otimes}_{i-1 \; \mbox{terms}} \hat{\sigma}^{+} \otimes \mathds{1} \otimes .. \otimes \mathds{1}, 
\end{equation}

In each cavity we can use the Fock basis $|0\rangle = [1, 0,0, ...,0]$, $|1\rangle = [0, 1, 0,..., 0]$, $|2\rangle = [0, 0, 1, ..., 0]$, and so on. In this basis, we have

\begin{equation} \label{a_boson}
\hat{a} = \left(\begin{array}{ccccc}
                   0 & \sqrt{1} & 0 & 0 & \ldots \\ 
				   0 & 0 & \sqrt{2} & 0 & \ldots \\ 
                   0 & 0 & 0 & \sqrt{3} & \ldots \\ 
                   0 & 0 & 0 & 0 & \ldots \\ 
                   \vdots & \vdots & \vdots & \vdots & \ddots \\ 
                   \end{array} \right).
\end{equation}

To construct a boson operator $\hat{a}$ in MATLAB we can write a = diag(sqrt(1:N)',1), where N is the maximum number of bosons and diag(X,1) returns a square upper diagonal matrix as show in equation~(\ref{a_boson}). The function acav() represent the operator $\hat{a}_i$ which is numerically defined as in Ref.~\cite{Korsch2016}, and is explicitly given by

\begin{lstlisting}[basicstyle=\tiny,style=Matlab-editor,basicstyle=\color{black}\ttfamily\scriptsize]
function x = acav(i,L,Nphoton,Is,Iatom)
a = diag(sqrt(1:Nphoton)',1);
a = kron(Iatom,a);
Op_total = cell(1,L);
for site = 1:L
    Op_total{site} = Is+double(eq(i,site))*(a-Is);
end
x = Op_total{1};
for site = 2:L
    x = kron(x,Op_total{site});
end
end
\end{lstlisting}

Similarly, the function sigmap() represent the operator $\hat{\sigma}_i^{+}$ and is given by

\begin{lstlisting}[basicstyle=\tiny,style=Matlab-editor,basicstyle=\color{black}\ttfamily\scriptsize]
function x = sigmap(i,L,Is,Icav)
up = [1 0]';
down = [0 1]';
sigma_p = up*down';
sigma_p = kron(sigma_p,Icav);
Op_total = cell(1,L);
for site = 1:L
    Op_total{site} = Is+double(eq(i,site))*(sigma_p-Is);
end
x = Op_total{1};
for site = 2:L
    x = kron(x,Op_total{site});
end
end
\end{lstlisting}

In the main code, particularly lines 46 and 48 we introduce the function QuantumSimulationCavityArray() which describes the construction of the Jaynes-Cummings-Hubbard and Rabi-Hubbard models for different detunings. Also, this functions introduce the many-body wave function, the time evolution, and the numerical calculation of the parameters given in equations (\ref{OrderParameter}) and (\ref{RateFunction}). The code for this function reads

\begin{lstlisting}[basicstyle=\tiny,style=Matlab-editor,basicstyle=\color{black}\ttfamily\scriptsize]
function [OP,P1m]= QuantumSimulationCavityArray(wc,D,g,Nphoton,L,A,Sp,N_ex,Hhopp,t,Model)
HJC = zeros((Nphoton+1)^L*2^L,(Nphoton+1)^L*2^L);
for i=1:L
    switch Model
        case 'Jaynes-Cummings'
            HJC = HJC + wc*A{i}'*A{i} + (D+wc)*Sp{i}*Sp{i}' + g*(Sp{i}*A{i}+Sp{i}'*A{i}');
        case 'Rabi'
            HJC = HJC + wc*A{i}'*A{i} + (D+wc)*Sp{i}*Sp{i}' + g*(Sp{i}+Sp{i}')*(A{i}+A{i}');
    end
end
H = HJC + Hhopp;        % Total Hamiltonian
dt = t(2)-t(1);         % Step time dt
U = expm(-1i*H*dt);     % Time propagator with step dt
up = [1 0]';            % Excited state for the two-level system
down = [0 1]';          % Ground state for the two-level system
Fock = eye(Nphoton+1);  % Fock states
theta1 = 0.5*atan(2*g*sqrt(1)/D);
phi_1m = cos(theta1)*kron(down,Fock(:,2))...
    -sin(theta1)*kron(up,Fock(:,1));  % Initial state |1,->
PSI_0 = phi_1m;
for k=1:L-1
    PSI_0 = kron(PSI_0,phi_1m);   % Many body wavefunction |1,->...|1,-> (L terms)
end
dn_T = zeros(size(t));            % Standard deviation dn_i = <n_i^2>-<n_i>^2
P1m = zeros(size(t));             % Population |1,-> at time t
for n=1:length(t)
    if n==1
        PSI = PSI_0;              % Initial wavefunction
    else
        PSI = U*PSI;              % Wavefunction at time t_n
    end
    dn = 0;
    for i=1:L
        dn = dn + PSI'*N_ex{i}^2*PSI-(PSI'*N_ex{i}*PSI)^2;
    end
    P1m(n) = abs(PSI_0'*PSI)^2;   % Ground state probability
    dn_T(n) = dn;                 % Standard deviation of the number of excitations
end
OP = mean(dn_T);  % Order parameter
end
\end{lstlisting}

\begin{figure}[htb]
\centerline{\includegraphics[width=1\textwidth]{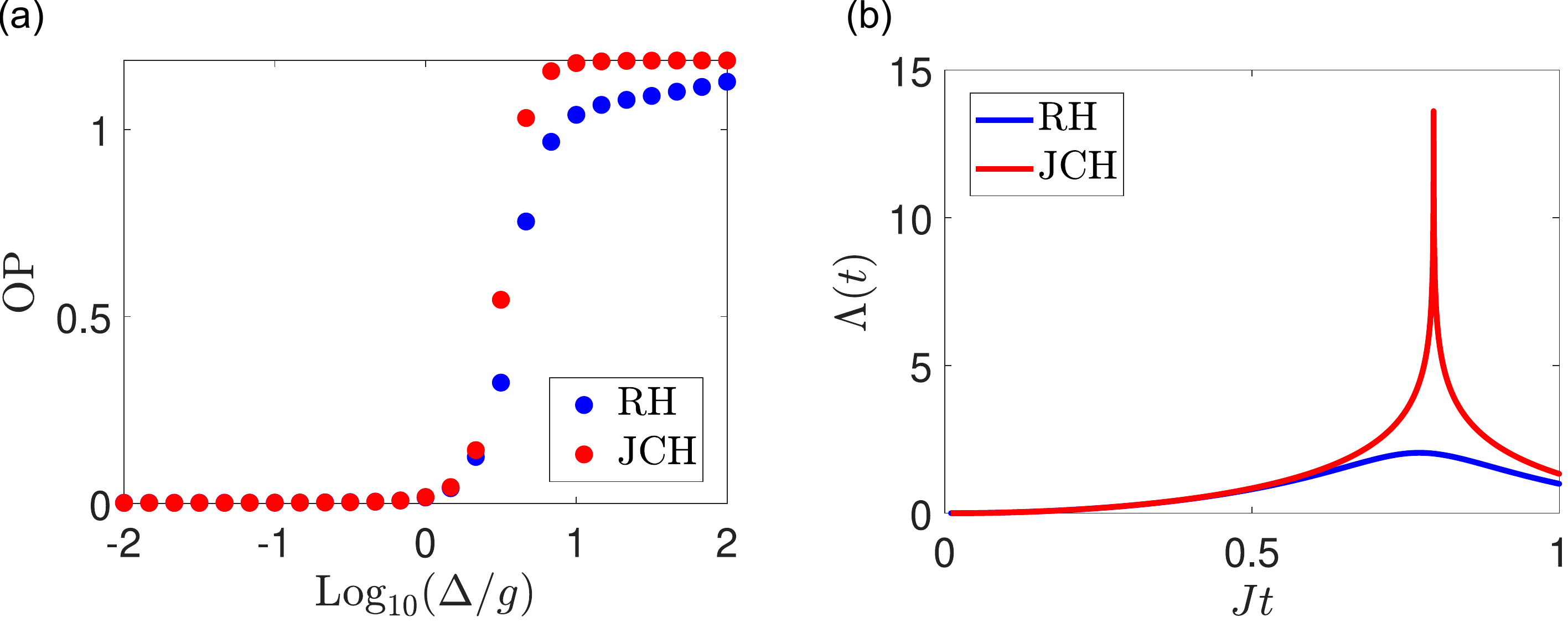}}
\caption{(a) Order parameter given in equation~(\ref{OrderParameter}) as function of $\mbox{log}_{10}(\Delta/g)$ for two interacting cavities using the Rabi-Hubbard (RH) and Jaynes-Cummings-Hubbard (JCH) models. We consider the values $g = 10^{-2}\omega_c$,  $J = 10^{-4}\omega_c$, and $N_{\rm cut-off} = 2$ as the cut-off parameter for the Hilbert space of each cavity. (b) Rate function $\Lambda(t)$ as a function of the dimensionless time $Jt$. This function exhibit non-analytic points at different times.} \label{figure1}
\end{figure}

The resulting order parameter~(\ref{OrderParameter}) and rate function~(\ref{RateFunction}) are plotted in figure~\ref{figure1} for two interacting cavities using the Jaynes-Cummings-Hubbard  and Rabi-Hubbard  models, respectively. From the numerical simulation we observe different curves for the order parameter in the region $\mbox{log}(\Delta/g)>0$. This is because the counter rotating terms $\hat{a}_i \hat{\sigma}_i^-$ and $\hat{a}_i^{\dagger}\hat{\sigma}_i^+$ (neglected in the RWA) are not negligible for $g = 10^{-2}\omega_c$. Furthermore, the rate function $\Lambda(t)$ has a remarkable non-analytic peak at $Jt \approx 0.8$, which is a characteristic signature of a dynamical quantum phase transition~\cite{Jurcevic2017, Wang2019}. The dynamical phase transition observed in the Jaynes-Cummings-Hubbard model (red lines in figure~\ref{figure1}-(b)) shows that the the probability to return to the Mott-insulator states, $P_{1-}(t) = |\langle \Psi(0) | \Psi(t) \rangle|^2$, is identically zero at the critical time $Jt \approx 0.8$. When $P_{1-}(t) = 0 $, and only for two cavities, the system is in the superfluid state, \textit{i.e.}, photons can freely move between cavities. \par

In the next section we will introduce a basic technique to address the numerical modelling of open quantum systems in the Markovian and non-Markovian regimes.

\section{Open quantum dynamics} \label{OpenQuantumSystems}

In this section, we introduce the Markovian and non-Markovian dynamics of open quantum systems. For the Markovian case, we will focus on the dynamical properties of the Lindblad master equation. We develop a fast and precise numerical method to solve the dynamics using the density matrix formalism. In the non-Markovian regime, we will examine the pure dephasing dynamics arising from the spin-boson model. We will explore memory effects by considering two different spectral density functions, say a super-ohmic and a Lorentzian models, and its effects on the time-dependent rate.

\subsection{Markovian quantum master equation}
Many open quantum systems can be modelled with a Markovian master equation in the weak coupling approximation~\cite{Breuer}

\begin{equation} \label{MarkovianMasterEquation}
{d \rho \over dt} = \mathcal{L}_{\rm M} \rho(t) = -i[\hat{H}_{\rm s},\rho(t)] + \sum_{i=1}^{N_{\rm c}}\gamma_i \left[\hat{L}_i \rho(t) \hat{L}_i^{\dagger} -{1 \over 2}\{\hat{L}_i^{\dagger}\hat{L}_i, \rho(t)\}\right], 
\end{equation}

where the first term in equation~(\ref{MarkovianMasterEquation}) is the conservative dynamics induced by the system Hamiltonian $\hat{H}_{\rm s}$. In contrast, the second term describes $N_{\rm c}$ dissipative channels through operators $\hat{L}_i$, that in the Markovian approximation can be associated with decay rates $\gamma_i>0$ for $i=1,...,N_{\rm c}$. The general solution of the presented master equation can be written 
as~\cite{Dominic2016,Katarzyna2016}

\begin{equation} \label{GeneralSolutionMarkovianMasterEquation}
\rho(t) = \sum_{k=1}^{N} c_k e^{\lambda_k t} R_k,
\end{equation}

where $c_k = \mbox{Tr}(\rho(0)L_k)$, $\lambda_k$ are the eigenvalues of the equation $\mathcal{L}_{\rm M}(R_k) = \lambda_k R_k$ and $\mathcal{L}_{\rm M}^{\dagger}(L_k) = \lambda_k L_k$, with $R_k$ and $L_k$ satisfying the orthonormality condition $\mbox{Tr}(R_k L_{k'}) = \delta_{kk'}$. The general solution given in equation~(\ref{GeneralSolutionMarkovianMasterEquation}) does not apply for time-dependent master equations. Therefore, the next method must be applied to systems described by a similar Lindblad structure as we showed in equation~(\ref{MarkovianMasterEquation}). To numerically solve the eigenmatrix equation $\mathcal{L}_{\rm M}(R_k) = \lambda_k R_k$ we adopt the formalism presented in Ref.~\cite{Carlos2015} to rewrite the Lindblad super-operator $\mathcal{L}_{\rm M} \rho(t)$. As an introductory example, we consider the open version of the transverse Ising model presented in Section~\ref{CloseQuantumSystems}

\begin{eqnarray}
{d \rho \over dt} &=& -i[-J S_1^{x} S_2^{x} - B(S_1^{x}+S_2^{x}),\rho(t)] \nonumber \\  
&&+ \gamma_1 \left[\hat{S}_1^{-} \rho(t) \hat{S}_1^{-,\dagger} -{1 \over 2}\{\hat{S}_1^{-,\dagger}\hat{S}_1^{-}, \rho(t)\}\right],  \nonumber \\
&&+ \gamma_2 \left[\hat{S}_2^{-} \rho(t) \hat{S}_2^{-,\dagger} -{1 \over 2}\{\hat{S}_2^{-,\dagger}\hat{S}_2^{-}, \rho(t)\}\right], 
\end{eqnarray}

where $S_{\alpha}^{-} = (\hat{S}_{\alpha}^x-i\hat{S}_{\alpha}^{y})/2$ for $\alpha = 1,2$ is the lowering spin operator. In the above equation $\gamma_i$ are associated with emission processes $\ket{\uparrow}_i \rightarrow \ket{\downarrow}_i$, where $\hat{S}_i^{z} \ket{\uparrow}_i = \ket{\uparrow}_i$ and $\hat{S}_i^{z} \ket{\downarrow}_i = -\ket{\downarrow}_i$. We proceed as follow, first we rewrite the density matrix of the $N$-dimensional system as a column vector~\cite{Carlos2015}

\begin{equation}
\vec{\rho}(t) = (\rho_{11}, \rho_{21},...,\rho_{N1},\rho_{12},...,\rho_{N2},...,\rho_{1N}, \rho_{2N}...,\rho_{NN})^{T},
 \end{equation}
 
In this vector representation, the master 
equation~(\ref{MarkovianMasterEquation}) takes the vector form $\dot{\vec{\rho}} = \mathds{L} \vec{\rho}$. The full matrix associated with the open evolution can be decomposed as $\mathds{L} = \mathds{L}_{H} + \mathds{L}_{\rm diss}$, where $\mathds{L}_{H}$ and $\mathds{L}_{\rm diss}$ account for the Hamiltonian and dissipative dynamics, respectively. First, $\mathds{L}_{H}$ can be coded as follow

\begin{lstlisting}[basicstyle=\tiny,style=Matlab-editor,basicstyle=\color{black}\ttfamily\scriptsize]
dim = 4;                  % Dimension of the total Hilbert space
Is = eye(dim);            % Identity matrix for the total Hilbert space
J = 1;                    % Value of the coupling term J
B = 0.1*J;                % Value of the magnetic field B
Sx = [0 1;1 0];    	  % S_x operator for one spin
Sz = [1 0;0 -1];          % S_z operator for one spin
I = eye(2);               % Identity matrix for one spin 1/2
H = -J*kron(Sx,Sx)-B*(kron(Sx,I)+kron(I,Sx)); % Hamiltonian of the system
L_H = -1i*kron(Is,H)+1i*kron(H.',Is);         % Lindblad operator L_H
\end{lstlisting}

By including the dissipative term $\mathds{L}_{\rm diss} = \sum_i \mathds{L}_i$, the total Lindblad operator can be written as

\begin{lstlisting}[basicstyle=\tiny,style=Matlab-editor,basicstyle=\color{black}\ttfamily\scriptsize]
gamma_1 = 0.1*B;       % Decay rate gamma_1
gamma_2 = 0.5*B;       % Decay rate gamma_2
S_minus =[0 0; 1 0];   % Lowering operator of the particle 1
L1 = kron(S_minus,I);  % Lowering operator of the particle 1 in the total Hilbert space
DL_1 = gamma_1*(kron(conj(L1),L1)-0.5*kron(Is,L1'*L1)-0.5*kron(L1.'*conj(L1),Is));
L2 = kron(I,S_minus);  % Lowering operator of the particle 2 in the total Hilbert space
DL_2 = gamma_2*(kron(conj(L2),L2)-0.5*kron(Is,L2'*L2)-0.5*kron(L2.'*conj(L2),Is));
L_diss = DL_1 + DL_2;  % Lindbladian L_diss
L = L_H + L_diss;      % Total Lindblad operator
\end{lstlisting}

\begin{figure}[htb]
\centerline{\includegraphics[width=1\textwidth]{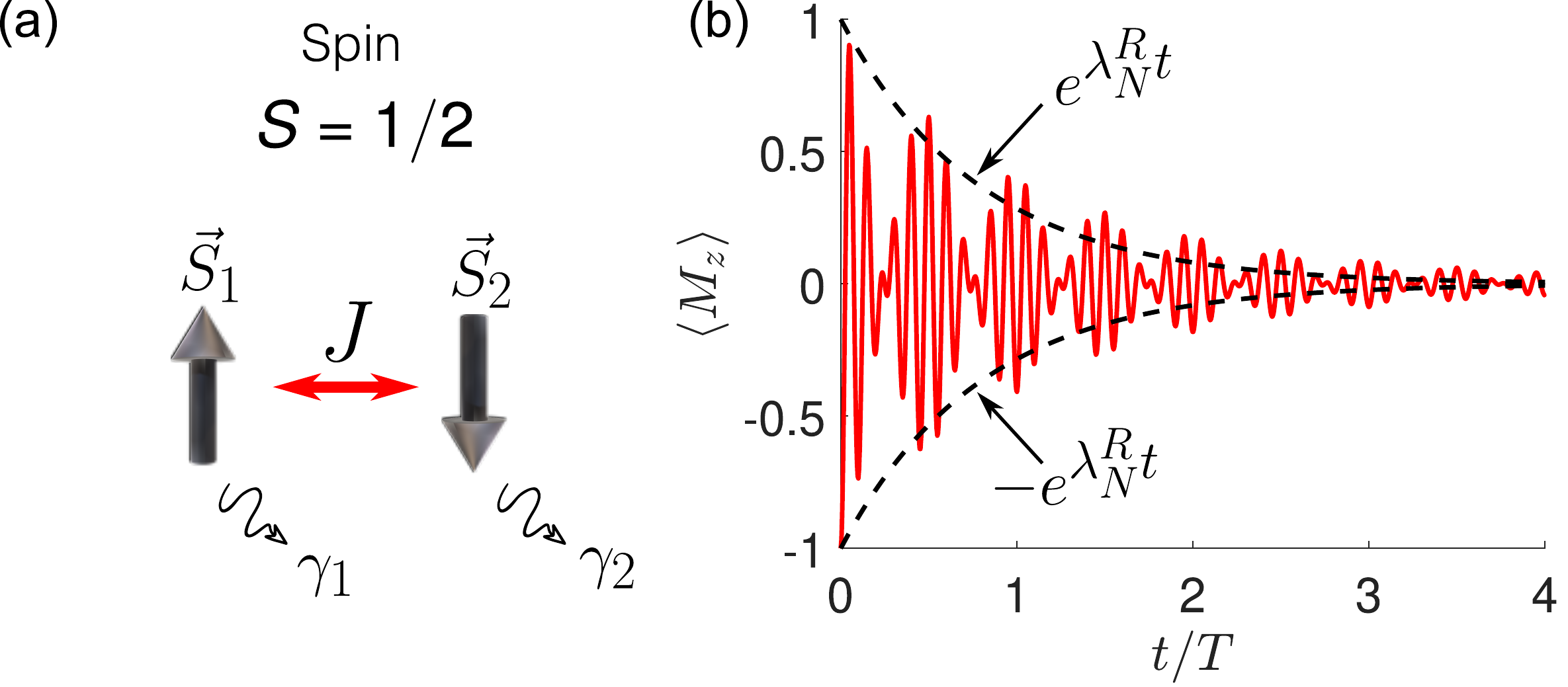}}
\caption{(a) Schematic representation of the dissipative Ising model. (b) Average magnetization along the $z$ direction for $J = 1$ and $B = J/10$ and considering the initial condition $\rho(0) = |\Psi(0)\rangle \langle \Psi(0)|$, with $|\Psi(0)\rangle = \ket{\downarrow}_1 \otimes \ket{\downarrow}_2$. The time is divided by the natural period $T = 2\pi/(2B)$. The envelope (dashed line) is calculated using the eigenvalue with the largest negative part $\lambda_N^R = -0.04$.} \label{example2}
\end{figure}

As the system consist in two interacting qubits, the density matrix $\rho(t)$ has a $4 \times 4$ size. As a consequence, $\vec{\rho}(t)$ has exactly $16$ elements. This implies that we have $16$ eigenvalues associated with $R_k$ and $L_k$ matrices. However, the eigenvalues and eigenmatrices must be sorted with the same criteria. To do this, we introduce the function sortingEigenvalues()

\begin{lstlisting}[basicstyle=\tiny,style=Matlab-editor,basicstyle=\color{black}\ttfamily\scriptsize]
function [R_sort,L_sort,lambda_sort] = sortingEigenvalues(dim,TOL,L)

[R,DR] = eig(L);       % Right eigenvectors and eigenvalues
[L,DL] = eig(L');      % Left eigenvectors and eigenvalues
eig_R = diag(DR);      % Right eigenvalues written as a vector
eig_L = diag(DL);      % Left eigenvalues written as a vector
ind_RL = zeros(dim*dim,2);
count = 1;
for n=1:dim*dim                 % Sorting of eigenvalues
    an = eig_R(n);
    for m=1:dim*dim
        bm = eig_L(m);
        if(abs(real(an)-real(bm))<TOL && abs(imag(an)-imag(bm))<TOL && count<=dim*dim)
            ind_RL(count,1) = n;
            ind_RL(count,2) = m;
            count = count + 1;
        end
    end
end
eig_L = eig_L(ind_RL(:,2)');    % Final sorting
eig_R = eig_R(ind_RL(:,1)');
L = L(:,ind_RL(:,2)');
R = R(:,ind_RL(:,1)');
lambda = eig_R;
[~,ind] = sort(lambda);         % Sorting of eigenvalues
lambda_sort = lambda(ind);           % \lambda_k eigenvalues
L = L(:,ind);
R = R(:,ind);

% Final R_sort and L_sort matrices
R_sort = cell(1,length(lambda_sort));
L_sort = cell(1,length(lambda_sort));
for k=1:length(lambda_sort)
    R_sort{k} = reshape(R(:,k),dim,dim);
    L_sort{k} = reshape(L(:,k),dim,dim);
    Rk = R_sort{k};
    Lk = L_sort{k};
    Ck = trace(Lk*Rk);
    Lk = Lk/sqrt(Ck);          % Normalized left eigenmatrices
    Rk = Rk/sqrt(Ck);          % Normalized right eigenmatrices
    R_sort{k} = Rk;
    L_sort{k} = Lk;
end
end
\end{lstlisting}

The above function returns the sorted eigenmatrices $R_k$, $L_k$ and eigenvalues $\lambda_k$ following a descending order for the real part of the eigenvalues: $0=\lambda_1^R \geq \lambda_2^{R} \geq ... \geq \lambda_{N-1}^{R} \geq \lambda_{N}^{R} $, where the $N$ eigenvalues are decomposed as $\lambda_k = \lambda^R_k + i\lambda_k^I$. The first eigenvalue $\lambda_1 = 0$ correspond to the steady state of the system, where $L_1 = \mathds{1}$~\cite{Dominic2016,Katarzyna2016}. Also, the eigenvalues with $k>1$ satisfy the condition $\lambda_k^{R}<0$ leading to dissipation terms $\propto e^{\lambda_k^{R}t}$ in the general solution defined in equation~(\ref{GeneralSolutionMarkovianMasterEquation}). The eigenvalue with the largest negative real part ($\lambda_{N}^{R}$) defines the envelope $e^{\lambda_N^R t}$ of the experimental observables (see figure~\ref{example1}-(b)). The normalized right ($R_k$) and left ($L_k$) eigenmatrices are calculated using the relations Rk = Rk/sqrt(Ck) and Lk = Lk/sqrt(Ck), where Ck = trace(Lk*Rk) is the normalization factor. Using the normalized matrices $R_k$ and $L_k$ we can compute any physical observable. By choosing the initial condition $\rho(0) = |\Psi(0)\rangle \langle \Psi(0)|$, with $|\Psi(0)\rangle = \ket{\downarrow}_1 \otimes \ket{\downarrow}_2$, we calculate the average magnetization $\langle  M_z \rangle = \mbox{Tr}(M_z \rho(t))$. The last part of the code read as

\begin{lstlisting}[basicstyle=\tiny,style=Matlab-editor,basicstyle=\color{black}\ttfamily\scriptsize]
TOL = 1e-10;
[R_sort,L_sort,lambda_sort] = sortingEigenvalues(dim,TOL,L);

down = [0 1]';                    % Quantum state down = [0 1]
Psi_0 = kron(down,down);          % Initial wavefunction
rho_0 = Psi_0*Psi_0';             % Initial density matrix
Nt = 3000;     	                  % Number of steps for time
T = 2*pi/(2*B) ;                  % Period of time
ti = 0;                           % Intial time
tf = 4*T;                         % Final time
dt = (tf-ti)/(Nt-1);              % Step for time
t = ti:dt:tf;                     % Time vector
Mz = zeros(size(t));              % Initial average magnetization
SSz = (kron(Sz,I)+kron(I,Sz))/2;  % Operator S1^z+S2^z
for n=1:length(t)        % Iteration to find general sulution of rho(t)
    rho = zeros(dim,dim);
    for k=1:length(lambda_sort)
        Lk = L_sort{k};
        Rk = R_sort{k};
        ck = trace(rho_0*Lk);
        rho = rho + ck*exp(lambda_sort(k)*t(n))*Rk; % General sulution for rho(t)
    end
    Mz(n) = trace(SSz*rho);   % General sulution for Mz(t)
end
figure()
hold on
plot(t/T,real(Mz),'r-','LineWidth',3)
plot(t/T, exp(real(lambda_sort(end))*t),'k--','LineWidth',2)
plot(t/T,-exp(real(lambda_sort(end))*t),'k--','LineWidth',2)
hold off
xlabel('$Bt$','Interpreter','LaTex','Fontsize', 30)
ylabel('$\langle M_z \rangle$','Interpreter','LaTex','Fontsize', 30)
set(gca,'fontsize',21)
\end{lstlisting}
 
The numerical stability of the presented method crucially depends on the value of the step size $dt$. The condition to have numerical stability is given by $dt \ll 1/\max_k(|\lambda_k|)$, where $\lambda_k$ are the eigenvalues associated to the equation $\mathcal{L}(R_k) = \lambda_k R_k$. In the above example we have chosen $dt = 0.0419$ since $\max_k(|\lambda_k|) = 2.2003$. For the rest of the examples we must ensure this condition.

In figure~\ref{example2} we plotted the expected average magnetization $\langle M_z \rangle$ for the dissipative two-spin system. In comparison with the non-dissipative case (see figure~\ref{example1}) the open Ising model shows a dissipative signal for $\langle M_z\rangle$. The envelope of this signal can be recognized as the exponential factor $\mbox{exp}(\lambda_N^{R}t)$ with $N = 16$ in our case. This dissipative behaviour is a consequence of the losses introduced in the Markovian master equation. \par

\subsection{Two-level system coupled to a photon reservoir}
 
In this subsection, we applied the previous algorithm to a different system, say, a two-level system interacting with thermal photons. We consider the following Markovian master equation for the atom-field interaction~\cite{Breuer}

\begin{eqnarray} \label{MarkovianMasterEquationTLSLight}
{d \rho \over dt} &=& {i\Omega \over 2}[\hat{\sigma}_+ + \hat{\sigma}_-,\rho(t)] + \gamma_0(N_{\rm ph}+1) \left[\hat{\sigma}_- \rho(t) \hat{\sigma}_-^{\dagger} -{1 \over 2}\{\hat{\sigma}_-^{\dagger}\hat{\sigma}_-, \rho(t)\}\right] \nonumber \\
  && + \gamma_0 N_{\rm ph} \left[\hat{\sigma}_+ \rho(t) \hat{\sigma}_+^{\dagger} -{1 \over 2}\{\hat{\sigma}_+^{\dagger}\hat{\sigma}_+, \rho(t)\}\right],
\end{eqnarray}

where $\Omega$ is the optical Rabi frequency, $N_{\rm ph}$ is the mean number of photons at thermal equilibrium, and $\hat{\sigma}_+ = |e\rangle \langle g| = \sigma_-^{\dagger}$. To solve the open dynamics of the reduced two-level system we implement the following code:

\begin{lstlisting}[basicstyle=\tiny,style=Matlab-editor,basicstyle=\color{black}\ttfamily\scriptsize]
Omega = 1;                      % Raby frequency from |1> to |2>
gamma0 = 0.2*Omega;             % Decay rate    
dim = 2;                        % Dimension Hilbert space two-level system
Is = eye(dim);                  % Identity matrix Hilbert space
v1 = Is(:,1);                   % Excited state for the atom
v2 = Is(:,2);                   % Ground state fro the atom
s11 = v1*v1';                   % Atom operator sigma_{11}
s22 = v2*v2';                   % Atom operator sigma_{22}
sp = v1*v2';                    % Atom operator sigma_{+}
sm = v2*v1';                    % Atom operator sigma_{-}
HL = -0.5*Omega*(sp+sm);        % Hamiltonian of the two-level system
N = 0;                          % Mean number of photons at zero temperature
Lrad = gamma0*(N+1)*(kron(conj(sm),sm)-0.5*kron(Is,sm'*sm)-0.5*kron(sm.'*conj(sm),Is)) + ...
       gamma0*N*(kron(conj(sp),sp)-0.5*kron(Is,sp'*sp)-0.5*kron(sp.'*conj(sp),Is));
L = -1i*kron(Is,HL)+1i*kron(HL.',Is)+Lrad;  % Lindblad operator
TOL = 1e-6;
[R_sort,L_sort,lambda_sort] = sortingEigenvalues(dim,TOL,L);
psi_0 = v2;                     % Ground state
rho_0 = psi_0*psi_0';           % Initial density matrix
Nt = 100000;                    % Number of steps for time
ti = 0;                         % Initial time
tf = 50/Omega;                  % Final time
dt = (tf-ti)/(Nt-1);            % Step time dt
t = ti:dt:tf;                   % Time vector
rho11 = zeros(size(t));         % Matrix elements \rho_{ij} = <i|\rho|j>
sigmap = zeros(size(t));
for n=1:Nt                      % General solution   
    rho = zeros(dim,dim);
    for k=1:length(lambda_sort)
        Lk = L_sort{k};
        Rk = R_sort{k};
        ck = trace(rho_0*Lk);
        rho = rho + ck*exp(lambda_sort(k)*t(n))*Rk;
    end
    rho11(n) = rho(1,1);
    sigmap(n) = trace(rho*sp);
end
% Exact Solution at zero temperature 
% p_e: population excited state, sp = <\sigma_x>
mu = 1i*sqrt((gamma0/4)^2-Omega^2);
pe_exact = Omega^2/(gamma0^2+2*Omega^2)*(1-exp(-3*gamma0*t/4).*(cos(mu*t)+3*gamma0/4/mu*sin(mu*t)));
if gamma0~=0
    sp_exact = -1i*Omega*gamma0/(gamma0^2+2*Omega^2)*(1-exp(-3*gamma0*t/4).*(cos(mu*t)+(gamma0/4/mu)-Omega^2/gamma0/mu*sin(mu*t)));
else
    sp_exact = -1i/2*sin(Omega*t);
end
figure
box on
hold on
plot(t*Omega,real(rho11),'r-','Linewidth',2)
plot(t*Omega,pe_exact,'b--','Linewidth',2)
hold off
xlabel('$\Omega t$','Interpreter','LaTex','Fontsize', 30)
ylabel('$p_e(t)$','Interpreter','LaTex','Fontsize', 30)
legend({'$\mbox{Numerical}$','$\mbox{Exact}$'},'Interpreter','latex','Fontsize', 21,'Location','northeast')
set(gca,'fontsize',21)
xlim([0 50])

figure
box on
hold on
plot(t*Omega,imag(sigmap),'r-','Linewidth',2)
plot(t*Omega,imag(sp_exact),'b--','Linewidth',2)
hold off
xlabel('$\Omega t$','Interpreter','LaTex','Fontsize', 30)
ylabel('$\mbox{Im}\langle \hat{\sigma}_+ \rangle$','Interpreter','LaTex','Fontsize', 30)
legend({'$\mbox{Numerical}$','$\mbox{Exact}$'},'Interpreter','latex','Fontsize', 21,'Location','northeast')
set(gca,'fontsize',21)
xlim([0 50])
\end{lstlisting}

The numerical method used to solve the Markovian dynamics of the two-level system coupled to photons is based on the general solution presented in equation~(\ref{GeneralSolutionMarkovianMasterEquation}). After define the Lindblad operator in line 15 (see source code), we find the eigenmatrices and eigenvalues using the same function sortingEigenvalues() presented in the previous section. The value of the tolerance parameter TOL (line 16 of the code) is chosen in order to have a sufficient numerical precision to discriminate the real and imaginary parts of the most similar eigenvalues $\lambda_k$.

\begin{figure}[htb]
\centerline{\includegraphics[width=1\textwidth]{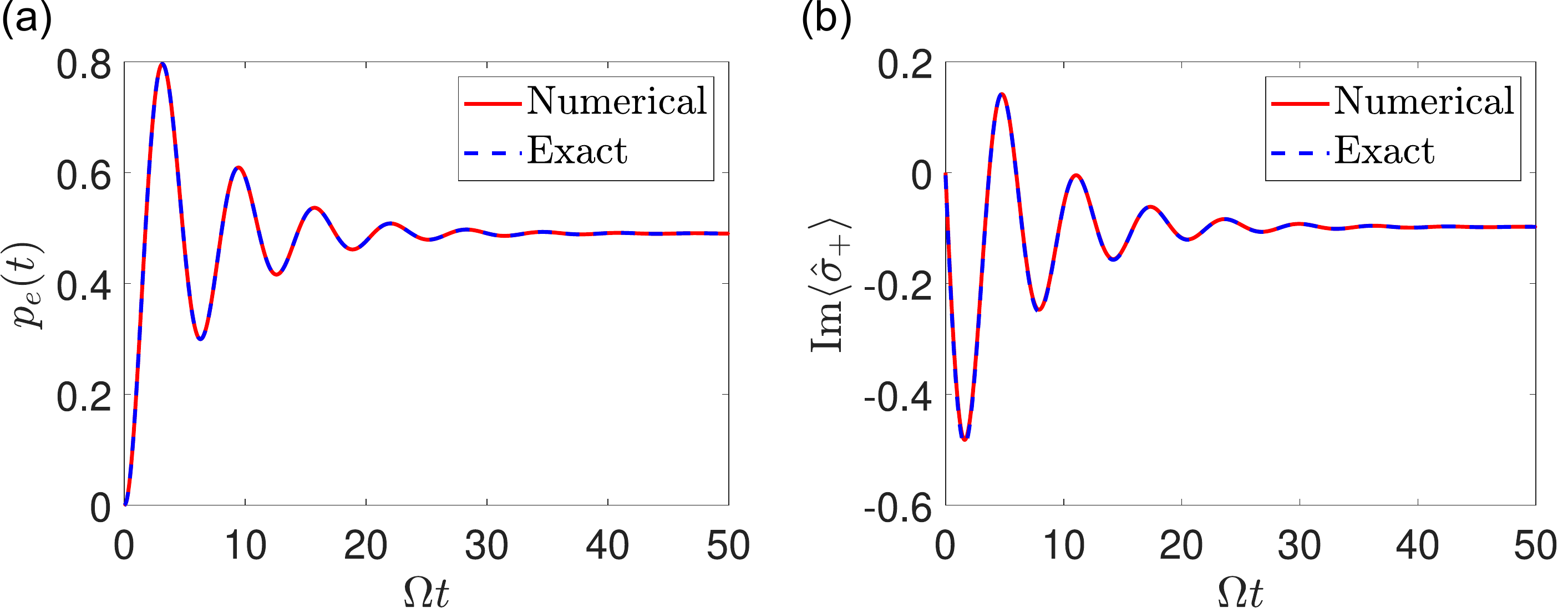}}
\caption{(a) Population of the excited state $p_e(t) = \langle e|\rho(t)|e\rangle$ for a two-level system interacting with a photon reservoir at zero temperature. In both curves we use $N=0$, $\gamma_0 = 0.2 \Omega$ and $\Omega = 1$. (b) Imaginary part of the observable $\langle \hat{\sigma}_+(t) \rangle$ as a function of time. In both plots the red (solid) and blue (dashed) lines correspond to the numerical and exact calculations, respectively.} \label{figure3}
\end{figure}

Following this procedure, in lines 30-33 of the source code, we calculated the density operator using the command rho = rho + ck*exp(lambda(k)*t(n))*Rk, where ck = trace(rho\_0*Lk), lambda(k) are the eigenvalues, Rk are the right eigenmatrices, and rho\_0 is the initial density matrix. \par

At zero temperature ($N_{\rm ph} = 0$) we have the following exact solutions 

\begin{eqnarray}
p_e(t) &=& {\Omega^2 \over \gamma_0^2+2\Omega^2} \left[1-e^{-3\gamma_0 t/4}\left(\cos \mu t + {3\gamma_0 \over 4\mu}\sin \mu t \right) \right], \\
\langle \hat{\sigma}_+(t) \rangle &=& {-i \Omega \gamma_0 \over \gamma_0^2+2\Omega^2}
\left[ 1-e^{-3\gamma_0 t /4}\left(\cos \mu t + \left\{ {\gamma_0 \over 4\mu}- {\Omega^2 \over \gamma_0 \mu} \right \} \sin \mu t\right)\right],
\end{eqnarray}

where $\mu = \sqrt{\Omega^2-(\gamma/4)^2}$. In figure~\ref{figure3} we plotted the population $p_e(t) = \langle e|\rho(t)|e\rangle$ and the imaginary part of $\langle \hat{\sigma}_+ \rangle$ for $N_{\rm ph}=0$.  We observed a good agreement between the numerical  and the exact solutions. Beyond the time evolution, the steady state of the system is a very useful information. For example, when the system reaches the steady state ($t\gg\gamma_0^{-1}$), the excited state and coherence can be found, $\rho_{ee}^{\rm ss} = \Omega^2 / (\gamma_0^2+2\Omega^2)$ and $\rho_{eg}^{\rm ss} =  i\Omega \gamma_0/(\gamma_0^2+2\Omega^2)$. In our case, $\Omega = 1$ and $\gamma_0 = 0.2 \Gamma$ resulting in $\rho_{ee}^{\rm ss}  = 0.4902$ and $\rho_{eg}^{\rm ss}  = 0.0980 i$. \par

The numerical approach to obtain the density matrix in the steady state reads

\begin{equation}
\rho_{\rm ss}^{\rm num} = c_1 R_1,
\end{equation}

This solution correspond to the particular case in which $\dot{\rho} = 0$. By looking our numerical simulations we obtain 

\begin{equation}
\rho_{\rm ss}^{\rm num}  = \left( \begin{array}{rr} 
                                                 0.4902 & 0.0980i \\
                               	               - 0.0980i  & 0.5098 
								\end{array} \right),
\end{equation}

Which exactly matches the theoretical predictions. Further extensions of this code to multiple level systems can be realized by changing the Hamiltonian and the dissipative contributions.

\subsection{Time-local quantum master equation}
In literature, the time-local quantum master equation in the secular approximation is presented as~\cite{Vega}

\begin{equation}
{d \rho \over dt} = \mathcal{L}_{\rm NM} \rho(t) = -i[\hat{H}_{\rm s},\rho(t)] + \sum_{i=1}^{N_{\rm c}}\gamma_i(t) \left[\hat{L}_i \rho(t) \hat{L}_i^{\dagger} -{1 \over 2}\{\hat{L}_i^{\dagger}\hat{L}_i, \rho(t)\}\right], 
\end{equation}

where $\gamma_i(t)$ are the time-dependent rates associated with operators $\hat{L}_i$. These rates can be obtained using a microscopic derivation ruled by the Hamiltonian describing the system-reservoir interaction~\cite{Breuer}. In order to numerically solve this type of equations we adopt a different approach with respect to the Markovian case. To solve the non-Markovian dynamics we employed a predictor corrector integrator method~\cite{NumericalRecipies,Korsch2016}. In the next section, we illustrate the main ideas by solving the pure-dephasing dynamics of the spin-boson model.

\subsection{Pure-dephasing model and non-Markovianity}
We consider the Hamiltonian for the pure dephasing spin-boson model ($\hbar = 1$)

\begin{equation} \label{PureDephasingHamiltonian}
H  = {\omega_{eg} \over 2} \hat{\sigma}_z + \sum_{k} \omega_k \hat{b}_{k}^{\dagger}\hat{b}_k + \frac{\hat{\sigma}_z}{2}\sum_{k} \left(g_k \hat{b}_k + g_k^{\ast}\hat{b}^{\dagger}_k \right), 
\end{equation}

where $\omega_{eg}$ is the bare frequency of the two-level system and $\omega_k$ are the boson frequencies. The exact time-local master equation in the interaction picture is~\cite{Luczka1990}

\begin{equation}\label{ExactPureDephasing}
{d \rho \over dt} =\frac{\gamma(t)}{2}\left[\hat{\sigma}_z\rho_s(t) \hat{\sigma}_z^{\dagger} - \{\hat{\sigma}_z^{\dagger} \hat{\sigma}_z, \rho(t)\}\right] =\frac{\gamma(t)}{2}\left[\hat{\sigma}_z\rho_s(t) \hat{\sigma}_z -\rho(t)\right].
\end{equation}    

The system-environment interaction is fully determined by the time-dependent dephasing rate ($\hbar = 1$)

\begin{equation} \label{DephasingRate}
\gamma(t) = \int_{0}^{\infty} {J(\omega) \over \omega} \coth\left({\omega \over 2 k_B T}\right)\sin(\omega t) \, d\omega,
\end{equation}

where $J(\omega) = \sum_{k} |g_k|^2\delta(\omega - \omega_k)$ is the spectral density function (SDF), $k_B$ is the Boltzmann constant and $T$ is the reservoir temperature. We solve the dynamics for the following spectral density functions

\begin{eqnarray}
J_1(\omega) &=& \alpha \omega_{\rm c}^{1-s} \omega^s e^{-\omega/\omega_{\rm c}}, \label{J1} \\
J_2(\omega) &=&  {J_0 \omega^s \over \left({\omega \over \omega_{0}}+1 \right)^{2}}\frac{\Gamma/2}{(\omega-\omega_{0})^2 + (\Gamma/2)^2},\label{J2} 
\end{eqnarray}

where $J_1(\omega)$ was originally introduced in the context of dissipative two-level systems~\cite{Leggett1987}. Physically, $\alpha$ is the system-environment coupling strength, $s \geq 0$ is a parameter, and $\omega_c$ is the cut-off frequency. Usually, three cases are defined: i) $0<s<1$ (sub-Ohmic), $s=1$ (Ohmic), and $s>1$ (super-Ohmic). The spectral density function $J_2(\omega)$ comes from the dynamics of quantum dots~\cite{WilsonRae2002} but also can describe localized-phonons in color centers in diamond~\cite{Ariel2019}.  In order to quantify the degree of non-Markovianity (NM) we use the following measure~\cite{Rivas2010,Rivas2014}

\begin{equation}\label{NM}
\mathcal{N}_{\gamma}(t) = {1 \over 2}\int_{0}^{t}(|\gamma^{\rm c}(\tau)|-\gamma^{\rm c}(\tau))\, d\tau,
\end{equation}

where $\gamma^{\rm c}(t)$ is the canonical rate when the master equation is written in the form $\dot{\rho} = \gamma^{\rm c}(t)\left[\hat{L}_z\rho_s(t) \hat{L}_z^{\dagger} - (1/2)\{\hat{L}_z^{\dagger} \hat{L}_z, \rho(t)\}\right]$ with $\mbox{Tr}(L_z^{\dagger}L_z) = 1$~\cite{Hall2014}. From equation~(\ref{ExactPureDephasing}) we recognize $L_z = \sigma_z/\sqrt{2}$ and  therefore we have $\gamma^{\rm c}(t) = \gamma(t)$. In what follows, we implement a code to numerically solve the time-dependent rate, the coherence, and the NM measure introduced in equation~(\ref{NM}).

\begin{lstlisting}[basicstyle=\tiny,style=Matlab-editor,basicstyle=\color{black}\ttfamily\scriptsize]
sz = [1 0;                            % Pauli matrix s_z
    0 -1];
Nw = 5000;                            % Number of points for \omega
wi = 0.01;                            % Initial frequency 
wf = 5;                               % Final frequency
dw = (wf-wi)/(Nw-1);                  % Step frequency d\omega
w = wi:dw:wf;                         % Frequency vector 
alpha = 5;                            % Strength spectral density function J1(w)
s = 2.5;                              % Ohmic-parameter s
wc = 0.1;                             % Cut-off frequency 
J1 = alpha*wc^(1-s)*w.^s.*exp(-w/wc); % Spectral density funcion J1(w)
J0 = 0.2;                             % Strength spectral density function J2(w)
w0 = 2;                               % Resonant frequency 
Gamma = 0.1;                          % Width of the spectral density function J2(w)
J2 = J0*(Gamma/2)./((w-w0).^2+(Gamma/2)^2).*w.^s./(w/w0+1).^2; % Spectral density funcion J2(w)
figure
box on
hold on
plot(w,J1,'r-','Linewidth',2)
plot(w,J2,'b-','Linewidth',2)
hold off
xlabel('$\omega$','Interpreter','LaTex','Fontsize', 30)
ylabel('$J(\omega)$','Interpreter','LaTex','Fontsize', 30)
legend({'$J_1(\omega)$','$J_2(\omega)$'},'Interpreter','latex','Fontsize', 21,'Location','northeast')
set(gca,'fontsize',21)
xlim([0 5])
Nt = 5000;                 % Number of points for time
ti = 0;                    % Initial time
tf = 100;                  % Final time
dt = (tf-ti)/(Nt-1);       % Step time dt
t = ti:dt:tf; t= t';       % Time vector
Psi_0 = [1 1]'/sqrt(2);    % Initial wavefunction
rho1 = Psi_0*Psi_0';       % Initial density matrix 
rho2 = rho1;
p11 = zeros(size(t)); 
p22 = zeros(size(t));
p12 = zeros(size(t)); 
p21 = zeros(size(t));
wa = ones(size(t))*w;      % Auxiliar frecuency vector
J1a = ones(size(t))*J1;    % Auxiliar J1 vector
J2a = ones(size(t))*J2;    % Auxiliar J2 vector
ta = t*ones(size(w));      % Auxiliar time vector
T = 0.001*w0;              % Temperature 
gamma1 = sum(J1a./wa.*sin(wa.*ta).*coth(wa/T/2),2)*dw;  % Rate gamma_1(t)
gamma_1_teo = alpha*wc*gamma(s)*sin(s*atan(wc*t))./(1+(wc*t).^2).^(s/2);
gamma2 = sum(J2a./wa.*sin(wa.*ta).*coth(wa/T/2),2)*dw;  % Rate gamma_2(t)
figure()
box on
hold on
plot(t,gamma1,'r-','Linewidth',2)
plot(t,gamma_1_teo,'k--','Linewidth',2)
plot(t,gamma2,'b-','Linewidth',2)
hold off
xlabel('$t$','Interpreter','LaTex','Fontsize', 30)
ylabel('$\gamma(t)$','Interpreter','LaTex','Fontsize', 30)
legend({'$\gamma_1(t)$','$\gamma_1^{\rm teo}(t)$','$\gamma_2(t)$'},'Interpreter','latex','Fontsize', 21,'Location','northeast')
set(gca,'fontsize',21)
xlim([0 100])
nk = 15;                  % nk steps per interval dt
C1 = zeros(size(t));
C2 = zeros(size(t));
N1 = zeros(size(t));
N2 = zeros(size(t));
for n = 1:length(t)
    for k = 1:nk
        L1 = gamma1(n)*(sz*rho1*sz -rho1);      % Lindbladian for gamma_1
        L2 = gamma2(n)*(sz*rho2*sz -rho2);      % Lindbladian for gamma_2
        rho1_pred = rho1 + L1*dt/nk;            % Predictor \rho_1
        rho2_pred = rho2 + L2*dt/nk;            % Predictor \rho_2
        rho1_m = 0.5*(rho1+rho1_pred);          % Mean \rho_1
        rho2_m = 0.5*(rho1+rho2_pred);          % Mean \rho_1
        L1 = gamma1(n)*(sz*rho1_m*sz -rho1_m);  % Lindbladian using mean \rho_1
        L2 = gamma2(n)*(sz*rho2_m*sz -rho2_m);  % Lindbladian using mean \rho_2
        rho1 = rho1 + L1*dt/nk;                 % Density matrix \rho_1
        rho2 = rho2 + L2*dt/nk;                 % Density matrix \rho_2
    end
    C1(n) = 2*abs(rho1(1,2));                   % Coherence C1(t)
    C2(n) = 2*abs(rho2(1,2));                   % Coherence C2(t)
    f1 = (abs(gamma1(1:n))-gamma1(1:n));   
    N1(n) = sum(f1)*dt;                         % Degree of non-Markovianity for gamma_1
    f2 = (abs(gamma2(1:n))-gamma2(1:n));
    N2(n) = sum(f2)*dt;                         % Degree of non-Markovianity for gamma_2
end
figure()
box on
hold on
plot(t,C1,'r-','Linewidth',2)
plot(t,C2,'b-','Linewidth',2)
hold off
xlabel('$t$','Interpreter','LaTex','Fontsize', 30)
ylabel('$C(t)$','Interpreter','LaTex','Fontsize', 30)
legend({'$C_1(t)$','$C_2(\omega)$'},'Interpreter','latex','Fontsize', 21,'Location','northeast')
set(gca,'fontsize',21)
xlim([0 50])

figure()
box on
hold on
plot(t,N1,'r-','Linewidth',2)
plot(t,N2,'b-','Linewidth',2)
hold off
xlabel('$t$','Interpreter','LaTex','Fontsize', 30)
ylabel('$\mathcal{N}_{\gamma}(t)$','Interpreter','LaTex','Fontsize', 30)
legend({'$\mathcal{N}_{\gamma_1}(t)$','$\mathcal{N}_{\gamma_2}(t)$'},'Interpreter','latex','Fontsize', 21,'Location','best')
set(gca,'fontsize',21)
xlim([0 100])
\end{lstlisting}

\begin{figure}[htb]
\centerline{\includegraphics[width=1\textwidth]{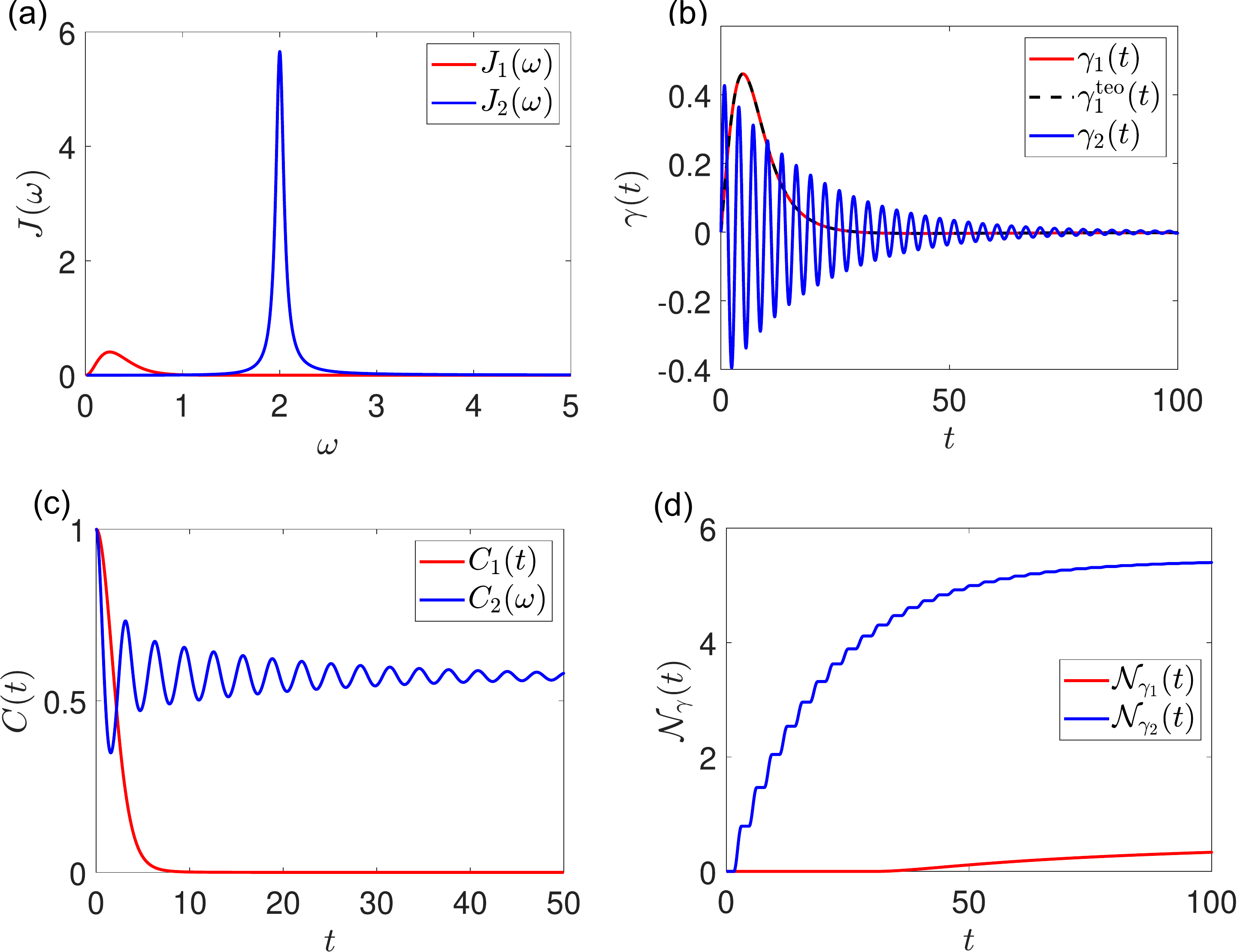}}
\caption{(a) Spectral density functions $J_1(\omega)$ and $J_2(\omega)$. For the spectral density functions we set the values $\alpha = 0.5$, $s = 2.5$, $\omega_{\rm c} = 0.1$, $J_0 = 0.2$, $\omega_0 = 2$, and $\Gamma = 0.1$. (b) Time-dependent rates $\gamma_1(t)$ and $\gamma_2(t)$ associated with $J_1(\omega)$ and $J_2(\omega)$, respectively. The rates are calculated at $T = 10^{-3}\omega_0$ (low-temperature). (c) Coherence function $C_1(t)$ and $C_2(t)$ associated with $\gamma_1(t)$ and $\gamma_2(t)$, respectively. (d) Degree of non-Markovianity $\mathcal{N}_{\gamma}(t) = (1/2)\int_{0}^{t}(|\gamma(\tau)|-\gamma(\tau))\, d\tau$.} \label{figure4}
\end{figure}

In figure~\ref{figure4}-(a) we plotted the spectral density functions $J_1(\omega)$ (red) and $J_2(\omega)$ (blue) given in equation~(\ref{J1}) and (\ref{J2}) for $s=2.5$ and $\omega_{\rm c} = 0.1$. The super-Ohmic function $J_1(\omega)$ reaches a maximum around $\omega \approx 0.25$ and quickly decreases due to the cut-off frequency term $\mbox{exp}(-\omega/\omega_{\rm c})$. On the contrary, the spectral density function $J_2(\omega)$ is strongly localized at the frequency $\omega_0 = 2$ with a full with at half maximum equal to $\Gamma = 0.1$. The associated rates $\gamma_1(t)$ and $\gamma_2(t)$ are illustrated in  figure~\ref{figure4}-(b). These time-dependent rates are calculated at low temperature $T = 10^{-3}\omega_0$. The damped oscillations of $\gamma_2(t)$ are a consequence of the strong interaction with the localized mode $\omega_0$. In fact, the periods of the signal are approximately given by $T \approx 2\pi /\omega_0 \approx 3.14$. On the other hand, the rate $\gamma_1(t)$ is positive in the time region $0 \leq t\leq 30.8$, while for $t>30.8$ the curve asymptotically reaches a constant negative value. In the low-temperature regime one theoretically obtains $\gamma_1^{\rm teo}(t) = \alpha \omega_{\rm c} \Gamma(s) \sin[s\tan^{-1}(\omega_{\rm c}t)]/[1+(\omega_{\rm c}t)^2]^{s/2}$~\cite{Maniscalco2013} with $\Gamma(s)$ being the gamma function (see dashed curve in figure~\ref{figure4}-(a)).  Thus, the negative region of $\gamma_1(t)$ is stablished by the condition $3\tan^{-1}(\omega_{\rm c}t)>\pi$ leading to the critical time $t_{\rm crit} = \tan(\pi/s)/\omega_{\rm c} \approx 30.8$, in good agreement with the numerical calculations. \par

The coherence functions $C(t) = \sum_{i\neq j}|\rho_{ij}(t)| = 2|\rho_{eg}(t)|$~\cite{Baumgratz2014} are shown in figure~\ref{figure4}-(c) for the initial condition $\rho(0) = |\Psi(0)\rangle \langle \Psi(0)|$ with $\Psi(0)\rangle = (|e\rangle + |g\rangle)/\sqrt{2}$. The super-Ohmic spectral density function $J_1(\omega)$ induces a monotonic decreasing behaviour in the coherence while the localized model introduces oscillations. These oscillations can be understood as a back-flow of quantum information between the system and the environment. Finally, the degree of NM $\mathcal{N}_{\gamma}(t)$ is calculated and shown in  figure~\ref{figure4}-(d). As expected, the localized model evidences a high degree of NM at any time in comparison with the super-Ohmic model. This can be  explained in terms of the fast oscillations observed in the rate $\gamma_2(t)$ and the coherence $C_2(t)$. Further extensions of this code can be easily done by assuming different terms in the Lindbladian and modifying the model for the environment. \par

\section{Concluding remarks}

In summary, we have developed selected examples to show how to code many-body dynamics in relevant quantum systems using the high-level matrix calculations in MATLAB. We oriented our discussion and examples to the fields of quantum optics and condensed matter, and we expect that the codes shown here will be valuable for graduate students and researchers that begin in these fields. Moreover, the simplicity of the codes will allow the reader to extend it to other similar problems. We consider that these codes can be used as a starting toolbox for research projects involving closed and open quantum systems.

\section{Acknowledgments}

The authors would like to thank Fernando Crespo for the valuable codes in Python for comparison. AN acknowledges financial support from Universidad Mayor through the Postdoctoral fellowship. RC and DT acknowledge financial support from Fondecyt Iniciaci\'on No. 11180143.

\section*{References}

\bibliographystyle{unsrt}

\end{document}